\documentclass[preprint2,floatfix,final]{aastex}

\usepackage{graphicx}
\usepackage[caption=false]{subfig}
\usepackage{psfrag}
\usepackage[fleqn]{amsmath}
\usepackage{mathtools}
\usepackage{natbib}
\usepackage{lscape}

\begin{document}

\title{ DYNAMICS OF DOUBLE LAYERS, ION ACCELERATION AND HEAT FLUX SUPPRESSION DU
RING SOLAR FLARES }


\author{T.C. Li\footnote{Present Address: Department of Physics and Astronomy, University of Iowa, Iowa City, IA 52242, USA. e-mail: takchu-li@uiowa.edu}, J.F. Drake and M. Swisdak } 
 
\affil{Institute for Research in Electronics and Applied Physics, University of Maryland, College Park, MD 20742, USA }



\begin{abstract}

Observations of flare-heated electrons in the corona typically suggest confinement of electrons. The confinement mechanism, however, remains unclear. The transport of coronal hot electrons into ambient plasma was recently investigated by particle-in-cell (PIC) simulations. Electron transport was significantly suppressed by the formation of a highly localized, nonlinear electrostatic electric potential in the form of a double layer (DL). In this work large-scale PIC simulations are performed to explore the dynamics of DLs in larger systems where, instead of a single DL, multiple DLs are generated. The primary DL accelerates return current electrons, resulting in high velocity electron beams that interact with ambient ions. This forms a Buneman unstable system that spawns more DLs. Trapping of heated return current electrons between multiple DLs strongly suppresses electron transport. DLs also accelerate ambient ions and produce strong ion flows over an extended region. This clarifies the mechanism by which hot electrons in the corona couple to and accelerate ions to form the solar wind. These new dynamics in larger systems reveal a more likely picture of DL development and their impact on the ambient plasma in the solar corona. They are applicable to the preparation for in-situ coronal space missions like the Solar Probe Plus.


\end{abstract}



\section{\label{intro}INTRODUCTION}


Novel advances in imaging and spectroscopy have made possible observations of solar eruptions such as flares and coronal mass ejections (CMEs) with unprecedented resolution \citep{Lin02}. This has led to better detection of radiation from energetic electrons in the tenuous corona, which, due to its lower plasma density, does not usually emit detectable radiation compared to the generally brighter emissions from the denser chromosphere. The corona is also proposed to be the electron acceleration site by flare models. Electrons can be accelerated to over 100 keV \citep{Lin03, Krucker10, Krucker07,Tomczak09}, three orders of magnitude higher than the ambient coronal temperature of $\sim$100 eV, during flares and CMEs. X-rays are emitted from the interaction between the accelerated electrons and ambient plasma. As energetic electrons propagate from the corona to the chromosphere, transport effects can modify the energy distribution of the propagating electrons and hence the observed X-ray spectra, affecting the interpretation of acceleration models. The transport of electrons from a coronal acceleration site is therefore crucial to understanding energy release in flares. 


Evidence for both confinement and free-propagation of energetic electrons is present in observations. A recent systematic study of solar flares with well-observed looptop (coronal) and footpoint (chromospheric) emissions found that the number of electrons required to explain observations is 2-8 times higher at the looptop than that at the footpoint \citep{Simoes13}. This implies electron accumulation at the looptop. Another systematic study of the relation between coronal and footpoint X-ray sources indicated that the simple scenario of electron free-streaming from the corona towards the footpoints could not explain the observed difference in photon spectral indices between the coronal and footpoint emissions \citep{Battaglia06}. A filter effect in the propagation preferentially reducing the distribution at lower energies was required. Such a filter can be an electric field. The observations of above-the-looptop sources reveal that the decay time of hard X-rays (HXR), which is a measurement of the lifetime of the HXR-producing electrons at the source, was over two orders of magnitude longer than the free-streaming transit time across the source \citep{Masuda94, Krucker07, Krucker10}. This suggests that the energized electrons are trapped at the looptop. A statistical survey of 55 partially occulted flares in which the bright footpoints were blocked by the solar limb showed that many looptop X-ray observations share similar decay times and sizes \citep{KruckerLin08}, implying electron confinement in the corona is a common phenomenon in solar flares.

On the other hand, there is evidence for free-propagation of energetic electrons, i.e., no interaction with the ambient plasma, in time-of-flight measurements of HXR emission \citep{Asch95,Aschwanden96,AschSchwartz95}. A systematic time delay between lower (25-50 keV) and higher (50-100 keV) energy HXR emission was reported from a statistical correlation study of over 600 flares \citep{Asch95}. This demonstrates that electrons with lower energies arrive at the chromosphere after those with higher energies as they freely stream down the flare loop from the corona. The measurements supported the free-streaming scenario, apparently contradicting the confinement scenario. An important component of solar flare models is the transport of energetic electrons from the corona to the chromosphere and from the corona to the solar wind. This, however, remains poorly understood.


The transport of electron energy has previously been modeled as both classical conduction \citep{Spitzer62} and anomalous conduction \citep{Manheimer77, Tsytovich71, Smith79, Levin93}. Recent analytical and numerical studies of the transport of super-hot electrons with energies greater than 10 keV showed that the classical conduction model produced a heat flux significantly higher than the real energy fluxes reported from multi-wavelength observations of solar flares \citep{Oreshina11}. This means that processes other than classical Coulomb collisions (between energetic electrons and ambient plasma) need to be considered as well.

In the anomalous conduction model, transport is limited by anomalous resistivity due to electron scattering in turbulent wave fields. Turbulence is produced by instabilities that result from the interaction between energetic electrons and ambient plasma. An anomalous conduction front at the head of an expanding hot electron source was considered as a way to confine hot electrons for the production of HXRs \citep{Smith79}. Since it was based on a one-fluid description, the model did not resolve the ion inertial length, let alone capture processes developing at electron scales. Later, 1D electrostatic particle-in-cell (PIC) simulations that resolved the shortest electron scale, the Debye length, did not reveal a conduction front \citep{McKean90}. Recently, the existence of a thermal conduction front was studied in 1D electrostatic Vlasov simulations \citep{Arber09}. They identified a propagating temperature jump as a conduction front. It was, however, suggested by recent PIC simulations and analysis \citep{Li13} that the propagating front was an ion acoustic shock and that the temperature jump was a result of shock heating at its extremely sharp transition. Other observed non-propagating temperature jumps associated with a potential jump were likely double layers (DLs), as discussed in Arber \& Melnikov (2009).

More recently, the transport of coronal energetic electrons was investigated by electromagnetic PIC simulations \citep{Li13}. It was shown that transport of flare-heated electrons from the source region was significantly suppressed. The suppression was due to the formation of a DL. Its associated electric potential reflected electrons back to the source region. A substantial fraction, about 50\%, of the total hot electron density was confined by the DL. There was also a population of escaping electrons, which may represent the free-streaming population discussed above. This model is in qualitative agreement with the observational evidence of confinement and free-propagation of energetic electrons.

A DL is a localized region that sustains a large potential drop in a collisionless plasma \citep{Block78,Raadu88}. The potential drop comes from a large-amplitude electrostatic electric field sandwiched between two adjacent layers of opposite charge. As a whole, the structure is neutral, but quasi-neutrality is locally violated within the layer, which occurs at scales of $\sim$10 Debye lengths $\lambda_{De}$. An ideal DL is a monopolar electric field, corresponding to a monotonic drop in the potential $\phi$. In general, a DL can be bipolar. There can be dips or bumps at the low or high potential sides, but an overall potential drop $\phi_{DL}$ develops across the structure. The strength of a DL is measured by $\phi_{DL}$. Particles are reflected by a DL if $\phi_{DL}$ is greater than their kinetic energy. For a thermal distribution, a DL with $\phi_{DL}$ equal to the temperature of the distribution can reflect the entire thermal bulk of the distribution, which contains particles with velocities lower than the thermal speed. In Li et al. (2013), it is the reflection of energetic electrons by a DL that confines them in the source region.

Although DLs have not yet been observed in the solar corona, they are widely seen in various space plasmas, including the solar wind \citep{Mangeney99}, the Earth's magnetosphere \citep{Mozer85, Ergun01, Ergun09} and Earth's radiation belt \citep{Mozer13}. They are also inferred to exist in the magnetospheres of Jupiter \citep{Hess09} and Saturn \citep{Gurnett12}. In the Earth's plasma sheet, a large number of DLs were detected by the THEMIS \citep{Angelopoulos08} spacecraft during periods of high magnetic activity, implying that DLs may frequently occur in such situations \citep{Ergun09}. In the solar atmosphere, many DLs are likely to develop during flares. Li et al. (2013) reported electron confinement by a single DL in PIC simulations. An important question is whether much larger, more realistic systems develop multiple DLs or a single dominant DL. If multiple DLs develop in larger scale systems, the transport properties of electrons are likely to be very different from that in a single DL system. Understanding this subject is important for future in-situ observations of the corona from missions like NASA's Solar Probe Plus \citep{Guo10}.

In this work, we investigate DL dynamics in simulation domains larger than those in earlier work. While only a single dominant DL is present in smaller domains, multiple DLs develop in larger domains. This gives rise to new dynamics not observable in a single-DL system. These dynamics are studied here for the first time. They include the following: the primary DL accelerates return current electrons to high velocities, producing electron beams that interact with ambient ions to drive the system Buneman unstable, resulting in the generation of many secondary DLs; trapping of electrons occurs between DLs, suppressing electron heat transport across these regions; DLs also accelerate ambient ions to produce strong ion flows across an extended region, which has important implications for understanding the formation of the solar wind. 


In the following, we describe the setup of the simulations and the parameters suitable for solar flare settings (Section \ref{sim}). The time evolution and key features of the simulations are summarized in Section \ref{evo_multi}. The generation of multiple DLs in larger domains are presented in Section \ref{mDL}. In Sections \ref{multi_gen} and \ref{ion_flow}, new dynamics involved in the production of secondary DLs and the interaction of these DLs with the ambient plasma are delineated. In Section \ref{dis6}, we briefly discuss the implications of the present results for understanding the mechanism for driving the solar wind, compare them with previous relevant studies and finally, we summarize our results. 

\section{\label{sim}SIMULATION SETUP}

\begin{figure}
\centering
\includegraphics[scale=0.3,trim=2.5cm 2.5cm 7cm 15.5cm, clip=true, totalheight=0.04\textheight]{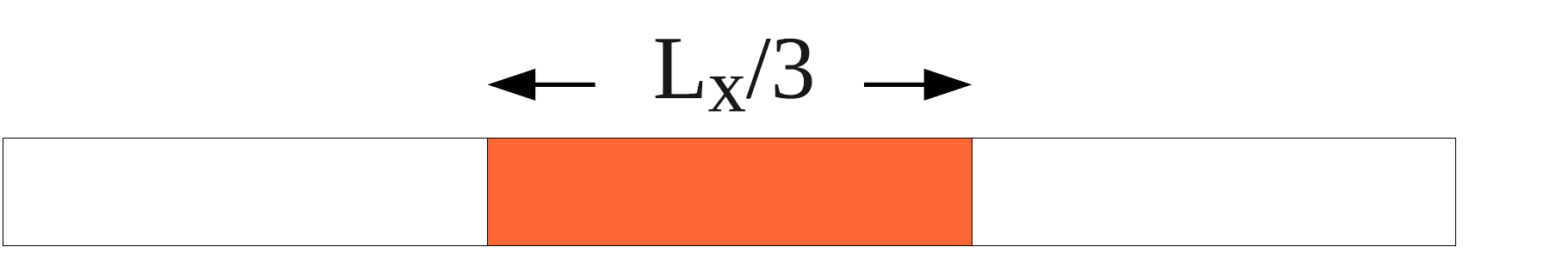}
\caption{ \label{box}  Schematic of the initial simulation setup. Adapted from Li et al. 2013.    }
\end{figure}

We perform two-dimensional electromagnetic PIC simulations using the p3d code \citep{Zeiler02}. The initial setup is the same as in Li et al. (2013). We use a rectangular domain (Figure \ref{box}) to represent a symmetric local segment of a flare loop centered at the looptop where a hot electron source is initialized. An initial uniform background magnetic field $B_0$ is applied in the long direction, $x$, which is along the loop axis. The initial density $n_0$ is uniform. Although the simulation domain is two-dimensional, no significant structure develops in the direction transverse to $B_0$.  Hence, we average over that coordinate to reduce noise. A third of the electrons, centered in the domain, have higher initial parallel temperature $T_{h0,\parallel}$ than the rest. It represents a pre-heated hot looptop source. Higher $T_{0,\parallel}$ is used because parallel transport dominates over perpendicular transport in the presence of a strong magnetic field, which is typical for a coronal loop. At the initial state, the parallel temperature transitions from a higher to a lower value over a length scale of a few tens of $\lambda_{De}$'s. The transition scale is not expected to affect the results because the temperature gradient broadens by more than an order of magnitude before DLs form. The broadening of the transition can be seen in Figure 2 of Li et al. (2012). In late time of the present simulations DLs form at various spatial locations far away from the the initial temperature gradient.

Electrons are initially modeled by bi-Maxwellian distributions and ions by a Maxwellian distribution. It is more natural that the hot population does not have a preferred direction of propagation. The simulated hot electrons are therefore not beamed in the initial state. We note that soft X-ray (SXR) counterparts of HXRs are observed in the high corona \citep{Masuda94,Krucker10}. SXR spectra can be well fit by a Maxwellian distribution \citep{Masuda00,Tsuneta97} and HXR spectra are usually fit to a combination of a thermal core and a nonthermal tail distribution. Although we use Maxwellian distributions in our simulation setup, we do not expect our results to be sensitive to the specific form of the initial electron distribution as long as there is a sufficient difference in the energy between the hot and cold regions. We use unity plasma beta $\beta$ (ratio of plasma pressure to magnetic pressure) in the parallel direction for hot electrons, corresponding to $T_{h0,\parallel}$=1. Using unity $\beta$ for hot electrons is consistent with recent coronal flare observations \citep{Krucker10}. Temperatures are normalized to $m_ic_A^2$, where $m_i$ is ion mass and $c_A$=$B_0/(4\pi m_in_0)^{1/2}$ is the Alfv$\acute{\mbox{e}}$n speed. Outside of the hot electron region, the ambient (cold) electron temperature $T_{c0}$ is 0.1, as are both the perpendicular temperatures throughout the domain and the ion temperature.

We perform four simulations with increasing domain sizes in the parallel direction, $L_x/L_{x0}$=1, 2, 4 and 8. $L_{x0}$ is the length of the smallest run, which has a size of $L_{x0} \times L_y$= 4634.1 $\times$ 18.1 $\lambda_{De}^2$. The grid size is 0.14$\times$0.14 $\lambda_{De}^2$. There are 400 particles per cell. $\lambda_{De}$=$v_{th0}/\omega_{pe}$ is the Debye length, $v_{th0}$=$(2T_{h0,\parallel}/m_e)^{1/2}$ is the thermal speed based on the initial hot parallel electron temperature and $\omega_{pe}$=$(4\pi n_0e^2/m_e)^{1/2}$ is the electron plasma frequency. Using typical parameters for coronal thermal X-ray sources, $T_h$=5 keV, n=10$^9$ cm$^{-3}$, we obtain $\lambda_{De}\sim$ 2 cm. The largest simulation thus has $L_x\sim$ 1 km, which is very small compared to the scale of a realistic flare loop. Space and time are normalized to $\lambda_{De}$ and $\omega_{pe}^{-1}$. A mass ratio $m_e/m_i$ of 1/100 and speed of light $c/v_{th0}$ of 7 are used. The system is periodic in both directions. Because of the periodic boundaries, the simulations are evolved for less than the electron transit time of the domain at 1.5$v_{h0}$, so the majority of hot electrons will not reach a boundary during a run. With increasing domain size, a simulation can be evolved for longer times, which allows us to study the long-time behavior of the system.


\section{\label{result}RESULTS}

\subsection{\label{evo_multi}The Impact of System Size on the Evolution of DLs}

\begin{figure*}[htb]
\centering
\includegraphics[scale=0.5]{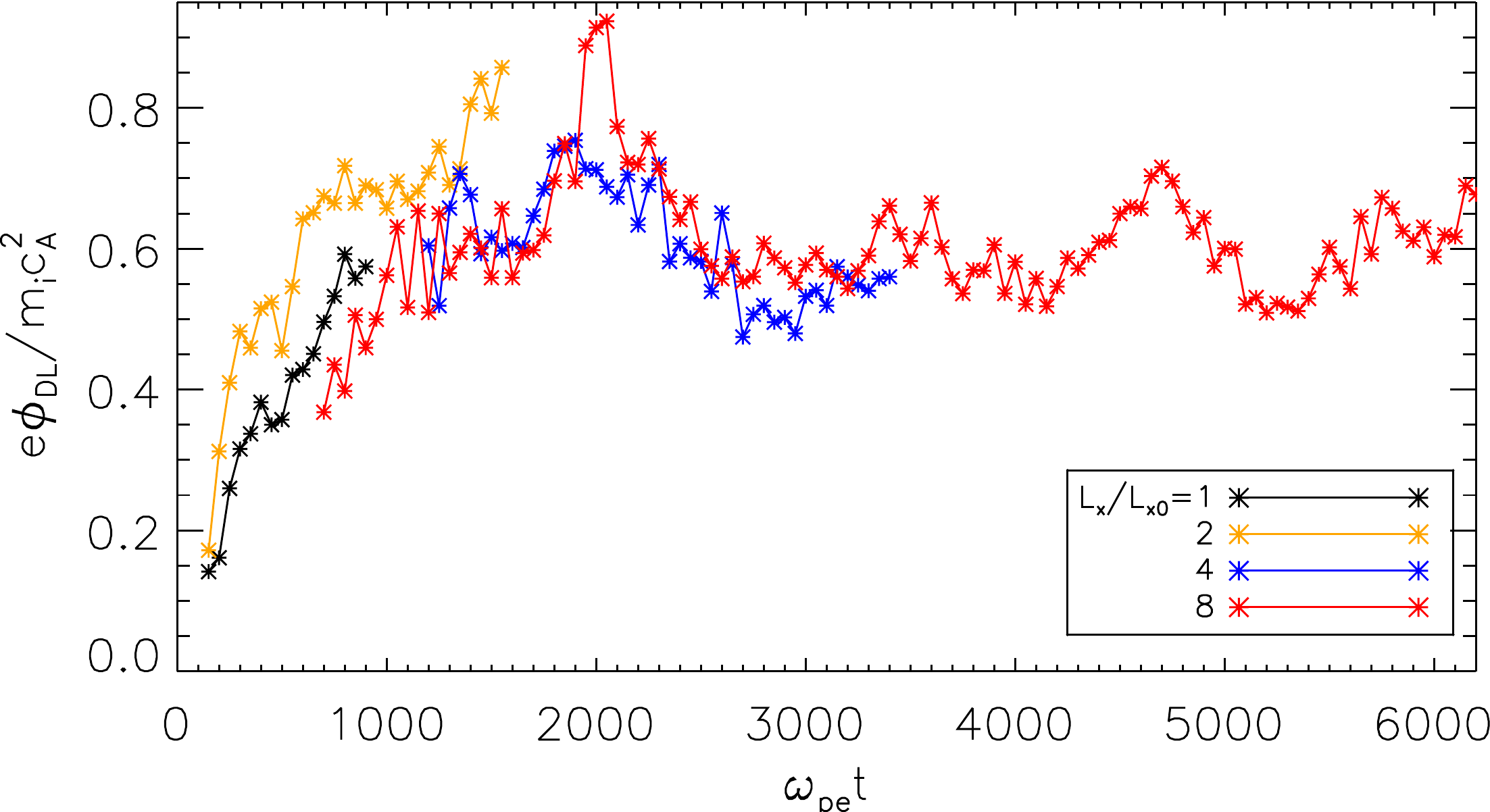}  
\caption{  \label{4phi} Evolution of the total DL strength $e\phi_{DL}$ in simulations with various domain sizes.  }
\end{figure*}

We present in figure \ref{4phi} the time evolution of the total DL strength $e\phi_{DL}$ in simulations with various system domain sizes. The largest domains contain more than one DL; for those the total DL strength is measured across all of them. $e\phi_{DL}$ is measured from the electric field data, which is averaged\footnote{The electric field data is averaged over one electron plasma period in the two smaller runs and over five periods in the two larger runs. Data is not sampled as frequently in the latter as in the former, leading to a longer averaging time and hence less effective subtraction of the fluctuations in the larger runs. $e\phi_{DL}$ of the smaller runs are therefore smoother while that of the largest run (red) appears to oscillate (about an overall growth trend) before $\omega_{pe}$t=1200, but in general becomes smoother after that time. No averaging is needed after 1200$\omega_{pe}^{-1}$ since the averaged and unaveraged data overlap. } to eliminate initial fluctuations at the contact of the hot and cold electron regions. The fluctuations fade out after the first $\sim$1200 $\omega_{pe}^{-1}$. $e\phi_{DL}$ at early times in the two larger runs are not shown because of the presence of oscillatory fluctuations that are not sufficiently averaged out due to the less frequent sampling of data, used for these runs to improve computational efficiency in larger systems.



The two smaller runs with $L_x/L_{x0}$=1 and 2 have similar evolutionary trends. The DL(s) in both cases grow, saturate and then grow again. The growth of DLs is driven by the Buneman instability due to the interaction of return current electrons and ambient ions \citep{Li12}. The saturation mechanism is the formation of a sound wave shock that develops from ions being accelerated by the DL potential to supersonic speeds \citep{Li13}. The shock stabilizes the Buneman instability and saturates further amplification of DLs. 

The larger simulation domains in the two largest runs allow us to follow the dynamics well beyond the growth phase of the DLs. At late time, $e\phi_{DL}/m_ic_A^2$ in the two largest simulations settles to an average value of $\sim$ 0.6. The DL strength is sustained throughout the entire course of the simulations without showing any signs of decay. In earlier 1D particle simulations in which DLs were driven by a strong applied potential, DLs decayed as many solitons, in the form of spiky wave trains, propagated toward the high potential side of the DLs \citep{Sato81}. They were observed to have a lifetime (of about 500 $\omega_{pe}^{-1}$) comparable to the transit time of the solitons across the DL width. In our case, the DLs remain for a much longer time and exhibit no sign of decay.

\subsection{\label{mDL}Formation of Mulitple DLs in Larger Systems}

\begin{figure}[tbp]
\centering
\includegraphics[scale=0.6]{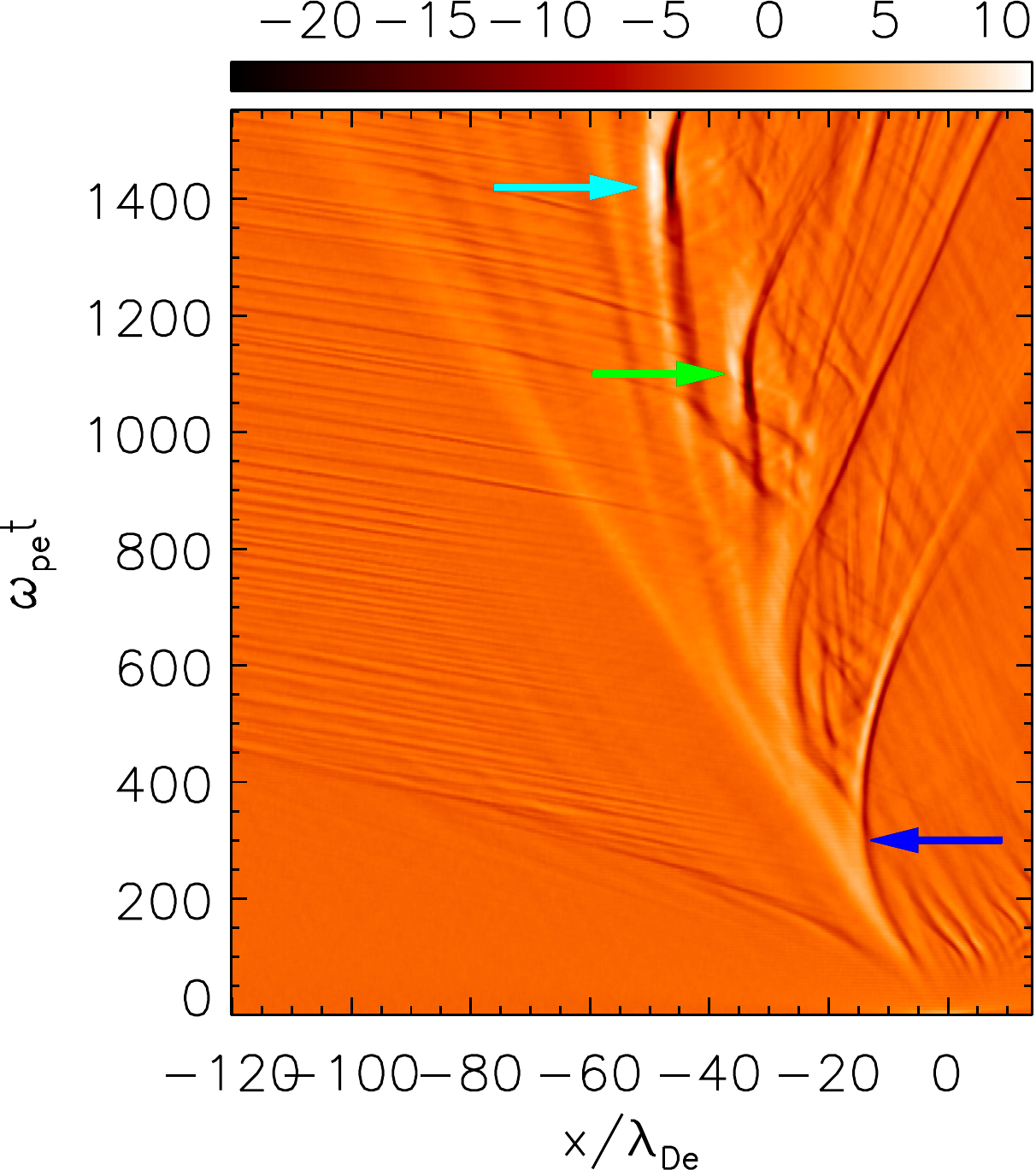}  
\caption{  \label{Ex_twoh} Time evolution of the electric field $E_x$ from the $L_x/L_{x0}$=2 run. The top two arrows indicate the position of two DLs and the bottom arrow locates a shock wave. }
\end{figure}

In the smallest run, a single dominant DL is observed (not shown). In the second run ($L_x/L_{x0}$=2), a weaker second DL emerges. We show in Figure \ref{Ex_twoh} the time evolution of the electric field $E_x$ from that run. The bright white feature indicated by a cyan arrow ($\omega_{pe}$t$\sim$1400) is the dominant DL. It sustains throughout the simulation. Another white feature indicated by a green arrow ($\omega_{pe}$t$\sim$1100) is the weaker second DL. A DL is signified by an overall jump in the electric potential $e\phi$, like those in Figure \ref{ep_early}(d) or Figure \ref{ph56}. A shock wave (blue arrow) is generated at $\omega_{pe}$t$\sim$300 (ion heating across the shock front in the form of ion trapping can be seen in Figure 6 of Li et al. (2013)). It leaves the DL at that time, causing saturation of the DL strength from $\omega_{pe}$=300 to 500. The plateauing of the orange curve in Figure \ref{4phi} indicates saturation during that period. We note that there is no significant jump in the electric potential at a shock after it leaves a DL although a DL and a shock appear similar in Figure \ref{Ex_twoh}. This will be demonstrated in the next figure.

In the two larger runs ($L_x/L_{x0}$=4, 8), many DLs are generated. In the following, we focus on data from the largest run to illustrate the dynamics in a multiple-DL system. To simplify notation, we drop the units of $x$ and $t$ from here on.

\begin{figure*}[tbh]
\centering
\includegraphics[scale=0.5]{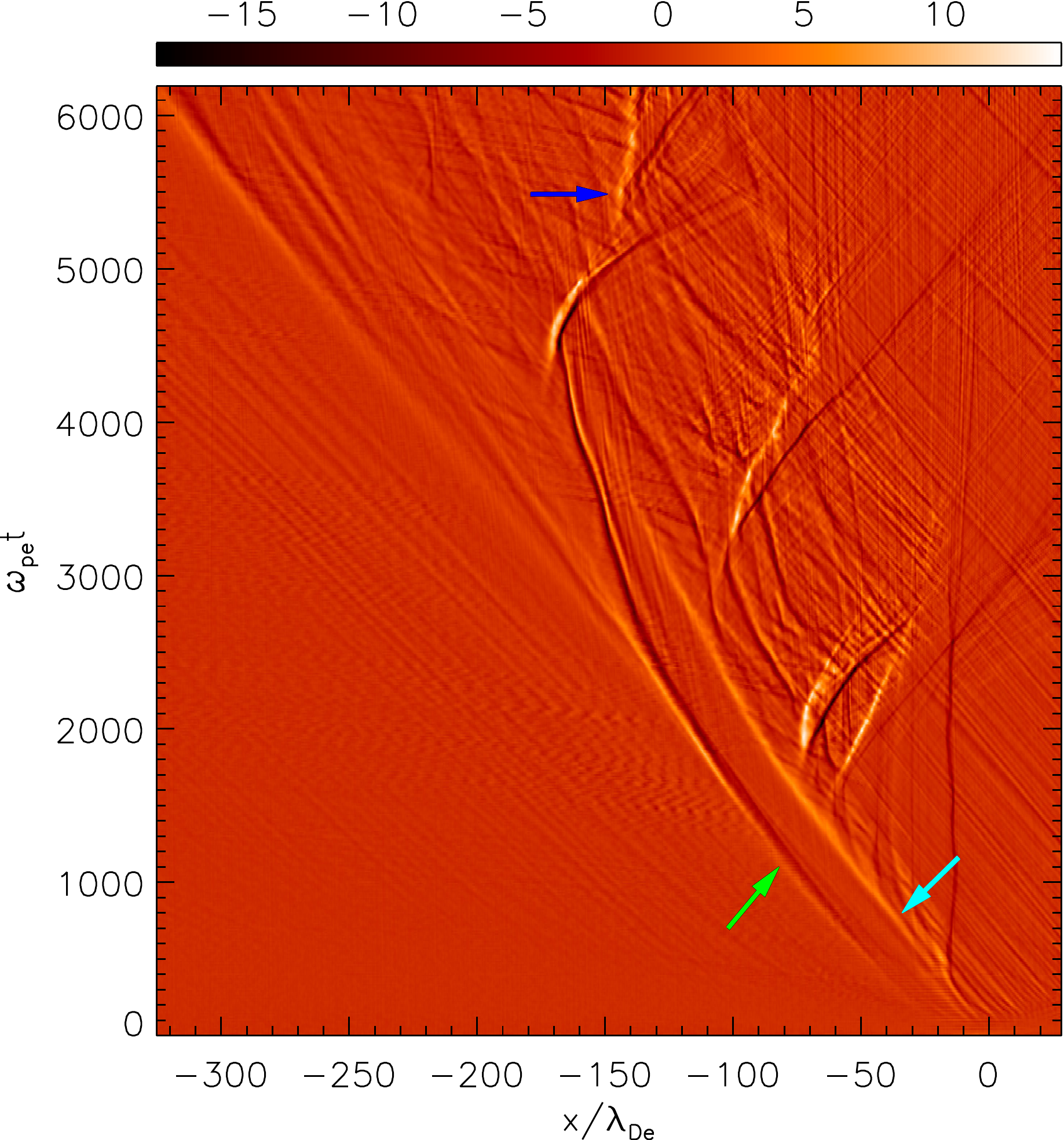}  
\caption{  \label{Ex_ehi} Time evolution of the electric field $E_x$ from the largest simulation. Several DLs are indicated by arrows. }
\end{figure*}

The evolution of $E_x$ from the largest simulation is shown in Figure \ref{Ex_ehi} . The first DL (cyan arrow) is the bright feature that emerges at $(x,t)\sim$ (0, 100), and is self-sustaining until $t\sim$ 3500 at $x\sim$ -140. It strengthens over time and propagates to the left. A shock associated with it forms at $(x,t)\sim$ (-14, 400) and separates from it thereafter. Note that at $(x,t)\sim$ (-15, 600), the shock, seen as a negative peak in $E_x$ in Figure \ref {ep_early} (d), does not cause a significant jump in $e\phi$ like a DL does. A second DL (green arrow) to the left of the first one emerges at $(x,t)\sim$ (-45, 400). A strong shock with an intense negative electric field is produced by this DL at $t\sim$ 2600 and separates from the DL. This will be further discussed later.

\begin{figure*}[htbp]
\centering
\subfloat(a){
\includegraphics[scale=0.55]{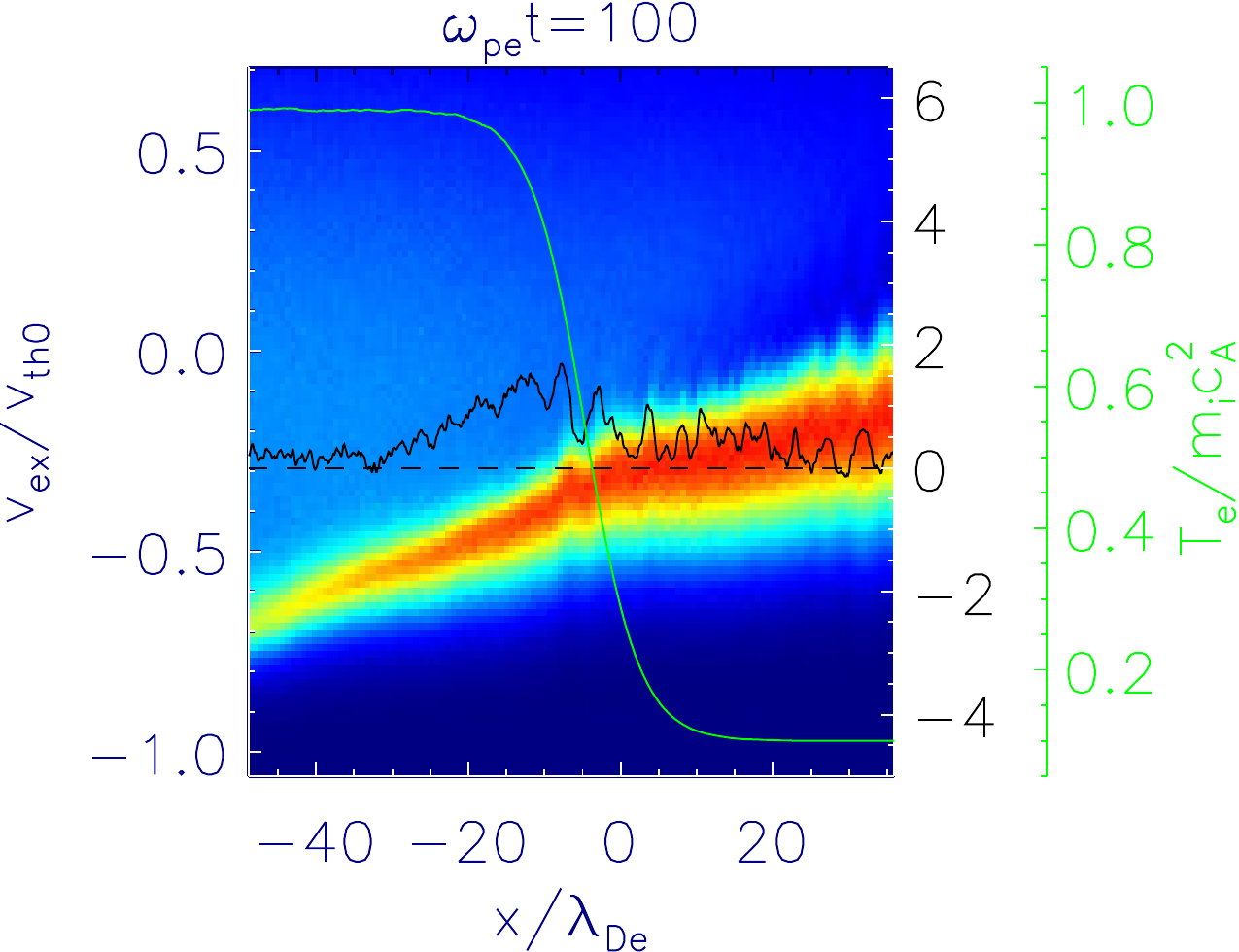}  
}\hspace{0.5em}%
\vspace{0.2em}%
\subfloat(b){
\includegraphics[scale=0.55]{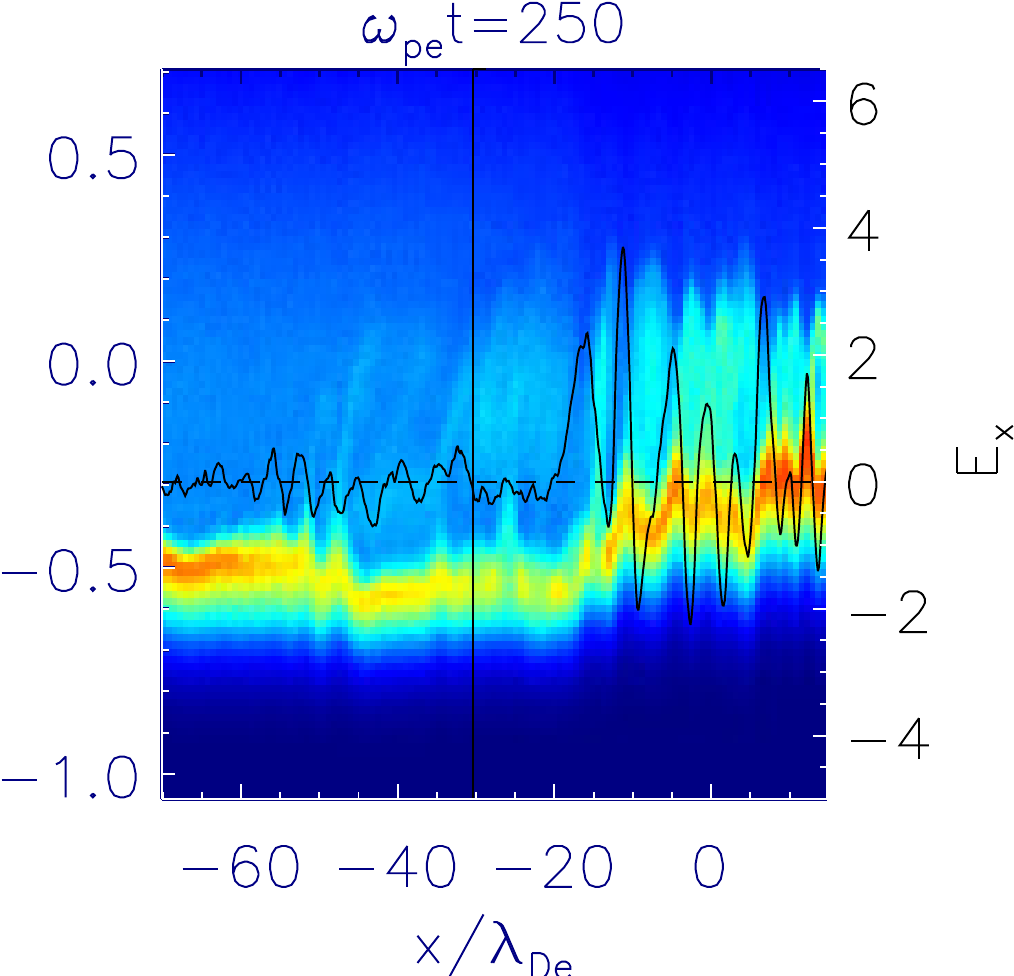}
}\hspace{2.5em}%
\vspace{0.2em}%
\\
\hspace{-2.5em}%
\subfloat(c){
\includegraphics[scale=0.55]{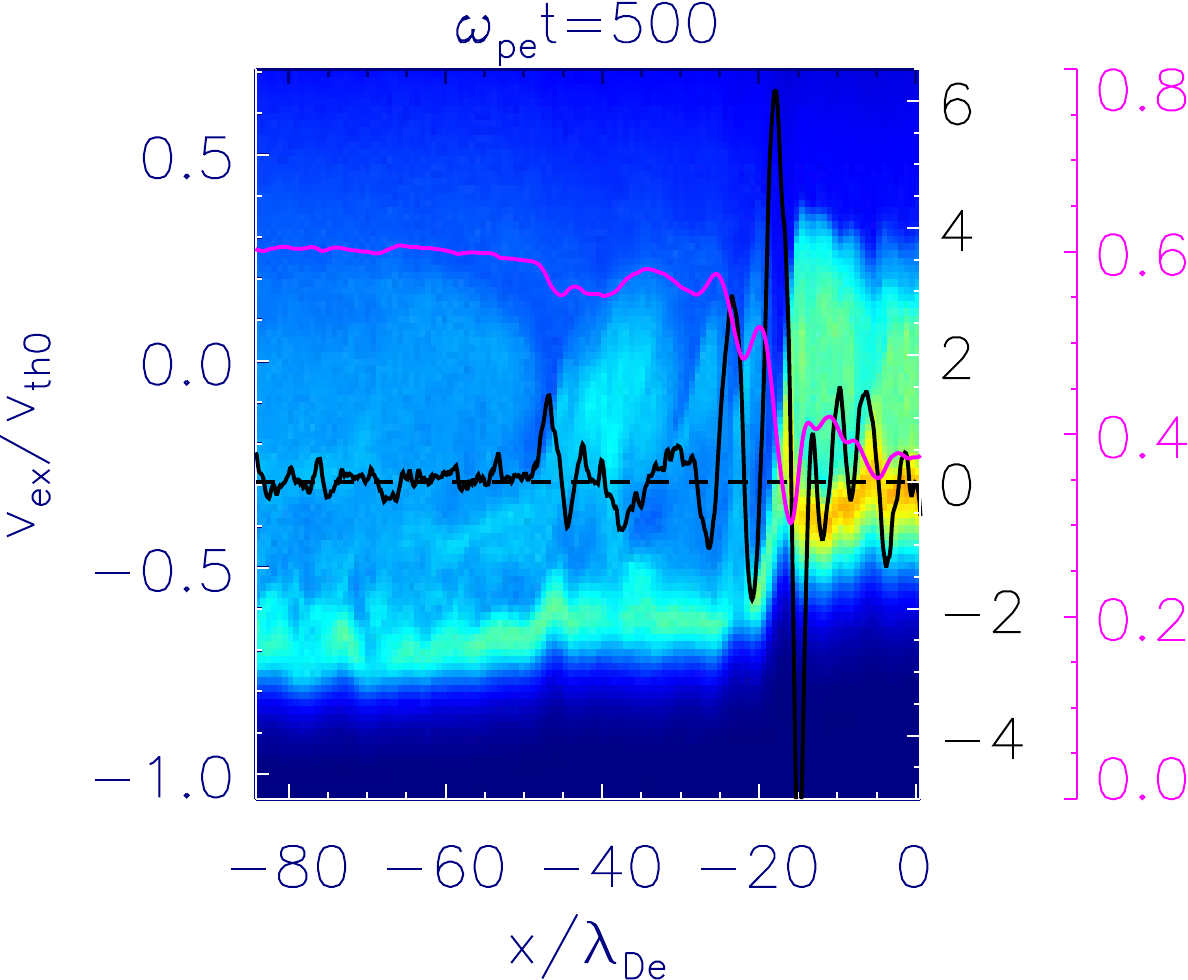} 
}
\vspace{-0.2em}%
\subfloat(d){
\includegraphics[scale=0.55]{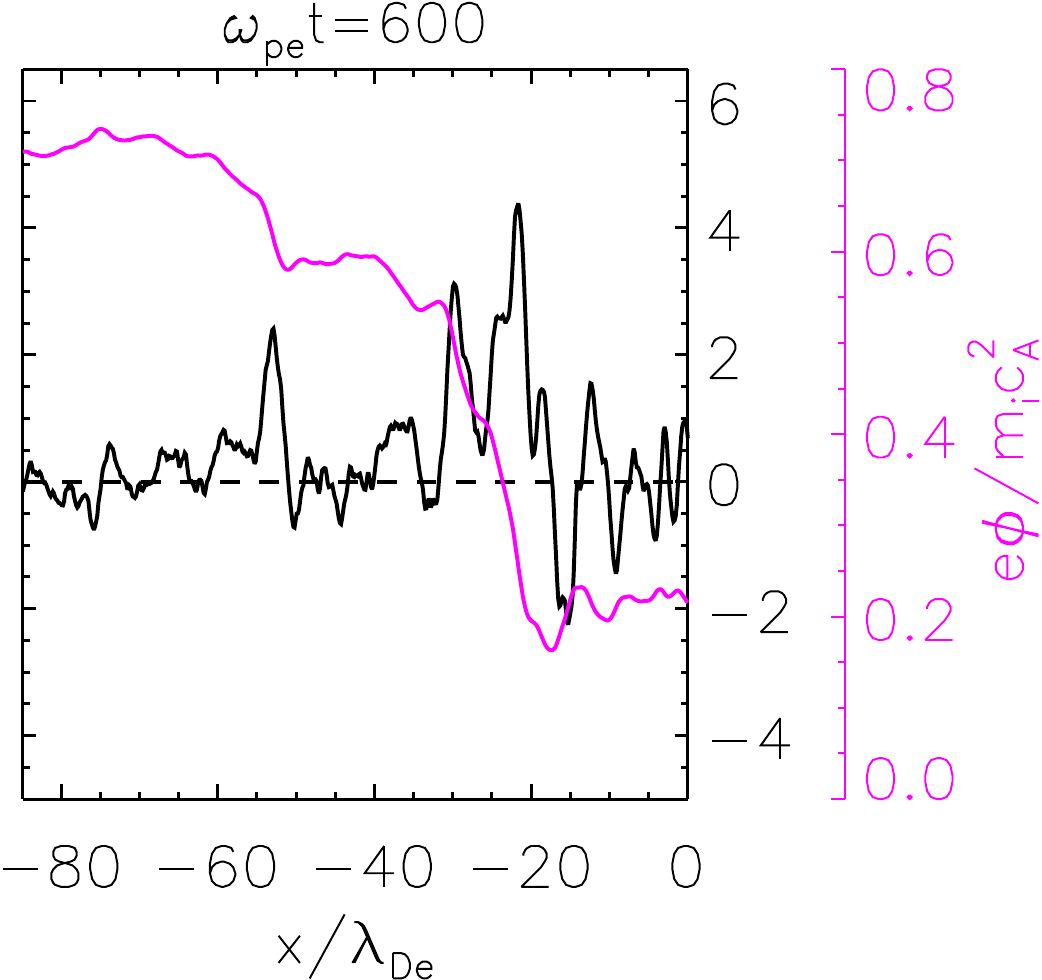}
}\hspace{3.em}%
\vspace{-0.2em}%
\centering
\caption{\label{ep_early} Electron phase space, during the emergence of the first and second DLs, (a) at t=100; (b) at t=250; and (c) at t=500. Velocities are normalized to the initial hot electron thermal speed $v_{th0}$=$(2T_{h0,\parallel}/m_e)^{1/2}$. The same color scale is used for all. Overlaid on (a)-(c) and plotted in (d) are $E_x$ (black), the electric potential $e\phi$ (magenta) and the electron temperature $T_e$ at t=0 (green).  }
\end{figure*}

\subsection{\label{multi_gen}Generation Dynamics in Multiple-DL Systems}

The fundamental generation mechanism for multiple DLs is the same as that for a single DL. As noted, it is due to the Buneman instability driven by the relative drift between a return current (RC) electron beam and the ambient ions \citep{Li12}. The RC beam comes from the cold electron region and is a response to the streaming hot electrons coming from the hot electron region. However, in a larger system, there is more time for the system to evolve and significantly different dynamics and interactions are observed. The original RC beam (that drives the first DL) is accelerated by the first DL and the resulting high-velocity beam is responsible for the generation of the second DL. Figure \ref{ep_early} illustrates the early time evolution of the electron phase space near the interface between the hot and (ambient) cold electrons. The initial (t=0) electron temperature (in green) is overlaid in (a) to show the hot and cold regions. To the right of $x\sim$0 is the cold electron region. To the left is the hot electron region. A RC beam that is drawn from the cold electrons into the hot region is the beam that excites the Buneman instability. On top of initial fluctuations in $E_x$ at $t$=100 (black), the first DL emerges among the unstable waves near $x\sim$ -10. The second DL is not yet present at this time. In (b), the first DL is developing at $x\sim$ -10 and the small second DL emerges at the position located by a vertical black line. The RC beam that is accelerated to $v_{ex}\approx$ -0.5 by the first DL is the driver of the second DL. The beam density (cyan) is being dragged from negative to positive velocities near the vertical black line, indicating partial reflection of the beam by the second DL. The electrons reflected to the right of the second DL, at $x\sim$ -45 in (c), increase from $t$=250 (b) to 500 (c), transferring energy to the second DL. As a result, the second DL strengthens noticeably from $t$=500 to 600 (d). This can be seen from the increase in the potential (magenta) drop across the second DL. Therefore, the first DL accelerates the RC beam and drives the formation of the second DL.

\begin{figure*}[tbh]
\centering
\includegraphics[scale=0.65]{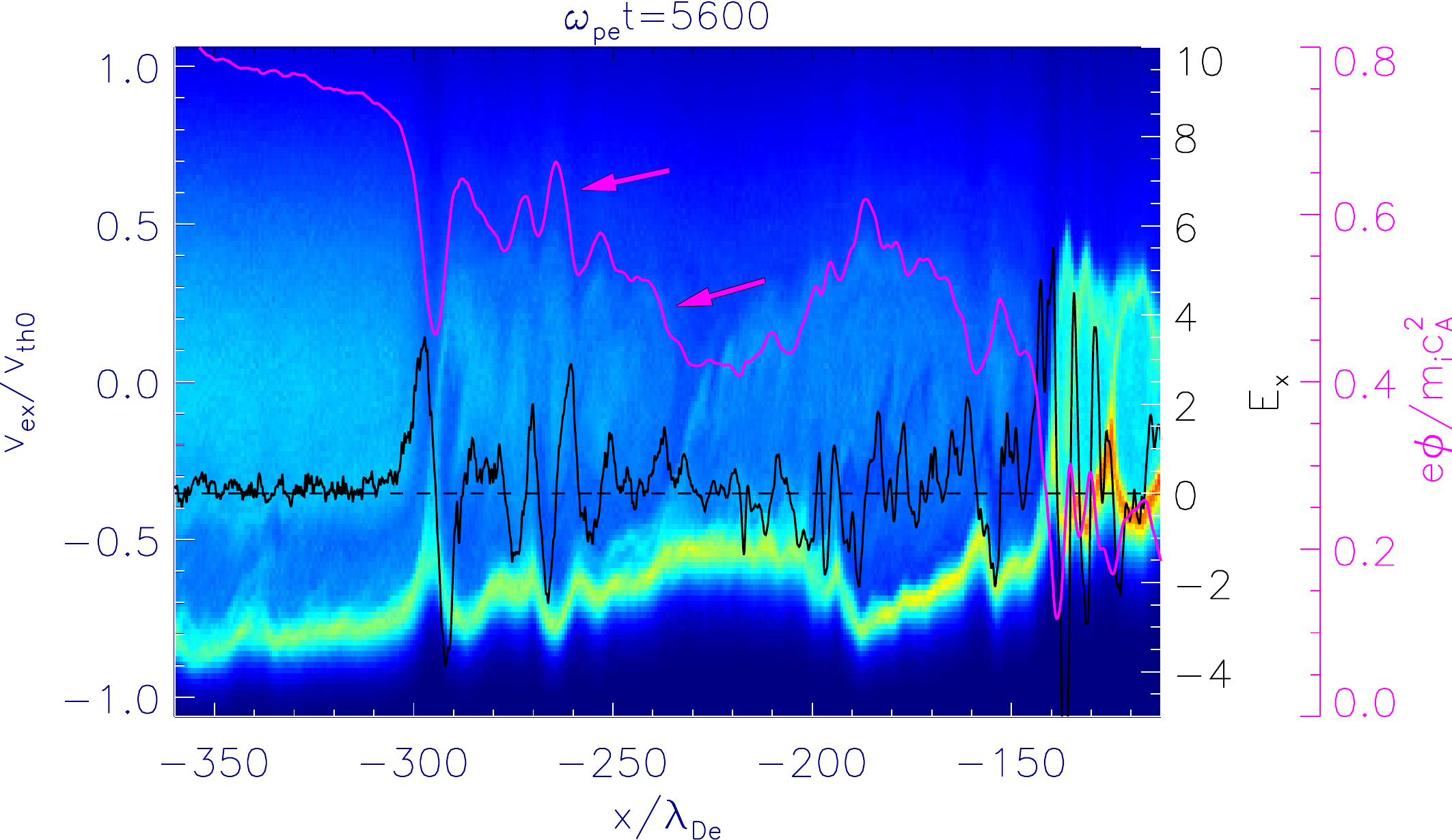} 
\caption{  \label{ph56} Electron phase space at $t$=5600, a time when many DLs form. Arrows indicate potential drops associated with a couple of small DLs between two larger DLs at x$\sim$ -300 and -143. The same format is used as in Figure \ref{ep_early}.  }
\end{figure*}


After a shock forms and departs at $t\sim$2600, the second DL weakens (Figure \ref{Ex_ehi}) because the shock reflects and hence holds back the RC electrons that can otherwise sustain the second DL \citep{Li13}. As the shock diminishes after $t\sim$5000, the second DL revives. In Figure \ref{Ex_ehi}, one sees a gradual brightening since $(x,t)\sim$ (-270, 5000). There is also increasing wave activity to the right of the second DL. Around $t\sim$ 5300, a third DL, indicated by a blue arrow in Figure \ref{Ex_ehi}, develops at $x\sim$ -145 and remains until the end of the simulation. It appears to propagate slowly to the right, opposite to what is usually observed. This is because the ions are accelerated by the second DL, producing an ambient flow to the right in the region $x>$ -270. As a result, the third DL is carried by the ambient plasma flow to the right.

In late time, the system reaches a state where many DLs arise. We show in Figure \ref{ph56} the electron phase space at $t$=5600, some time after the emergence of the third DL. In addition to the second DL (at $x\sim$ -300) and the third DL ($x\sim$ -140), where significant potential jumps occur, a couple of weaker DLs are generated and overall there is substantial wave activity. Magenta arrows indicate the associated potential jumps. Note that the RC beam is reflected by the negative $E_x$ of the second DL and also by the weaker DLs (for instance, around $x\sim$ -235).

In larger domains that allow the system to evolve for a much longer time, the behavior at late time is much more dynamic than what was seen in smaller domains (Li et al. 2013, Sato \& Okuda 1981). Multiple DLs are generated downstream of the RC beam from a primary DL (the first DL in this case). The growth of these downstream (to the left of the primary DL) DLs produces an ambient ion flow that pushes the upstream DL further to the right, opening up a space (in between) for more DLs to form. A chain of many DLs results.

\subsection{\label{ion_flow}Interaction with Ambient Ions}

\begin{figure*}[tbh]
\centering
\includegraphics[scale=0.5]{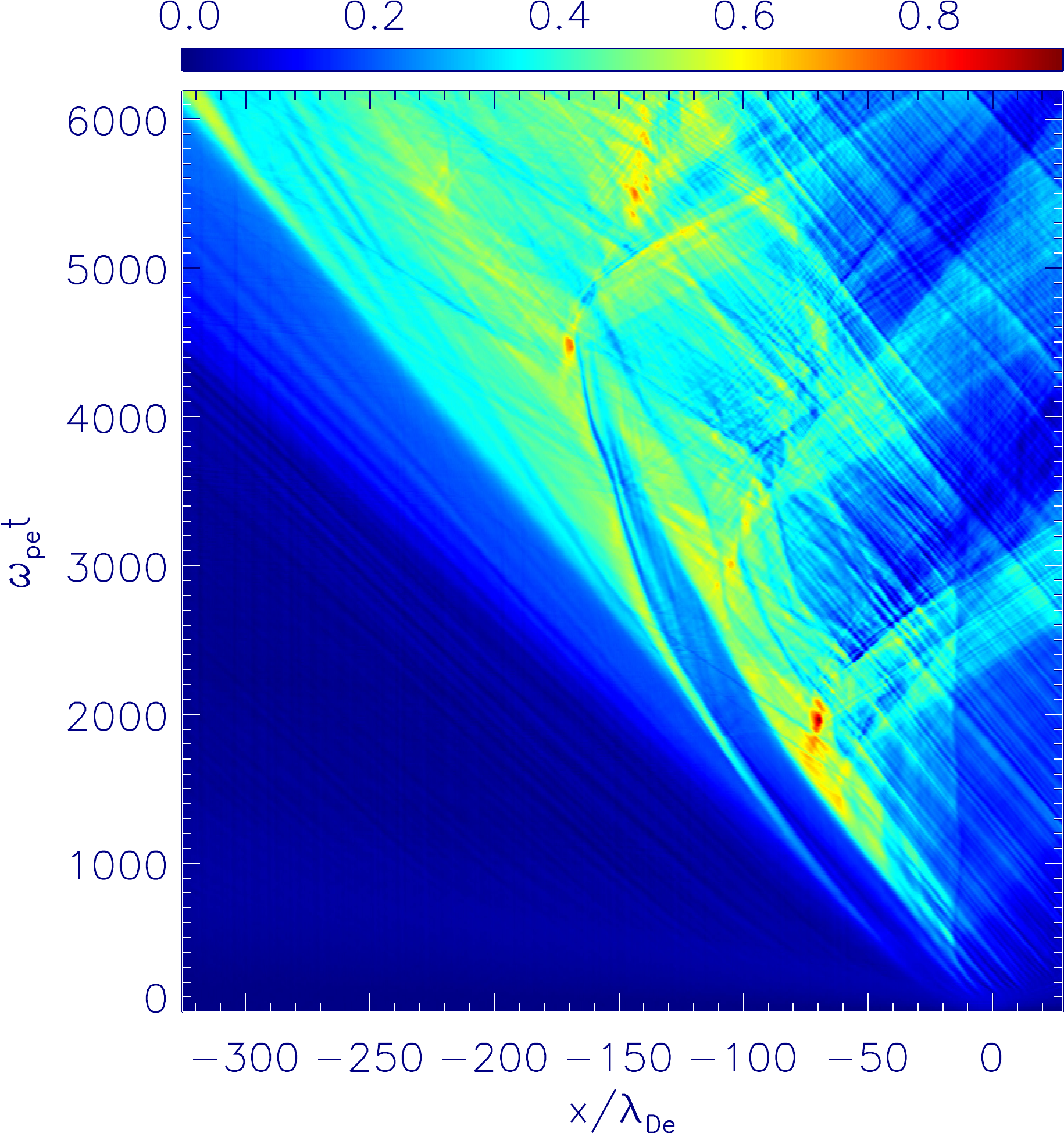}  
\caption{  \label{svix_ehi} Time evolution of the ion flow velocity $v_{ix}/c_A$ in $x$, analogous to the electric field evolution in in Figure \ref{Ex_ehi}.}
\end{figure*}

As discussed, DLs interact with ambient ions and create strong ion flows in a multi-DL system. We present in Figure \ref{svix_ehi} the space-time evolution of the ion flow velocity. When compared to the electric field evolution in Figure \ref{Ex_ehi}, one can see that the ions are strongly accelerated at the locations of DLs. The change in ion flow therefore resembles the pattern of the spatial and temporal development of the electric field. Following the path of the first DL, significant flow velocities of $\sim$0.4 $c_A\sim$1 $v_{ti}$(=$\sqrt{2T_i/m_i}$) are produced.  After being accelerated by the DLs to velocities as high as $\sim$2 $v_{ti}$ at the red patches in Figure \ref{svix_ehi}, ions flow to the right (due to a drop in $e\phi$) and the flow extends over a wider scale than the acceleration site, which is localized at the DLs. This is essentially the ambient plasma flow that carries the third DL to the right as mentioned in Section \ref{multi_gen}. Over time, as DLs spawn across the system, away from the proximity of the initial contact between the hot and cold electron regions,  strong ion flows spread out in space as well. We discuss the implication of this process in Section \ref{dis6}.

\begin{figure*}[htb] 
\centering
\subfloat(a){
\includegraphics[scale=0.43]{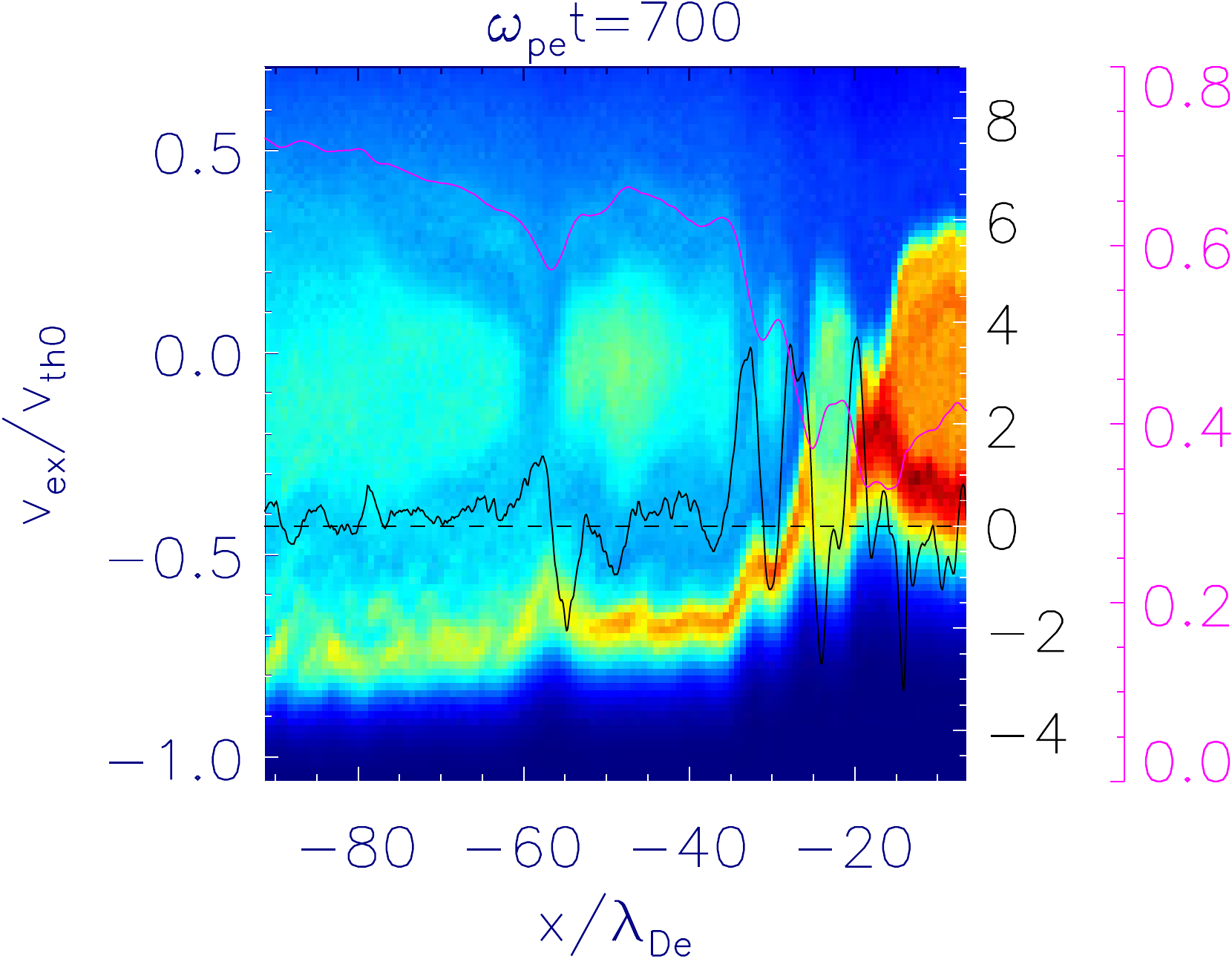}   
}
\vspace{0.3em}%
\subfloat(b){
\includegraphics[scale=0.43]{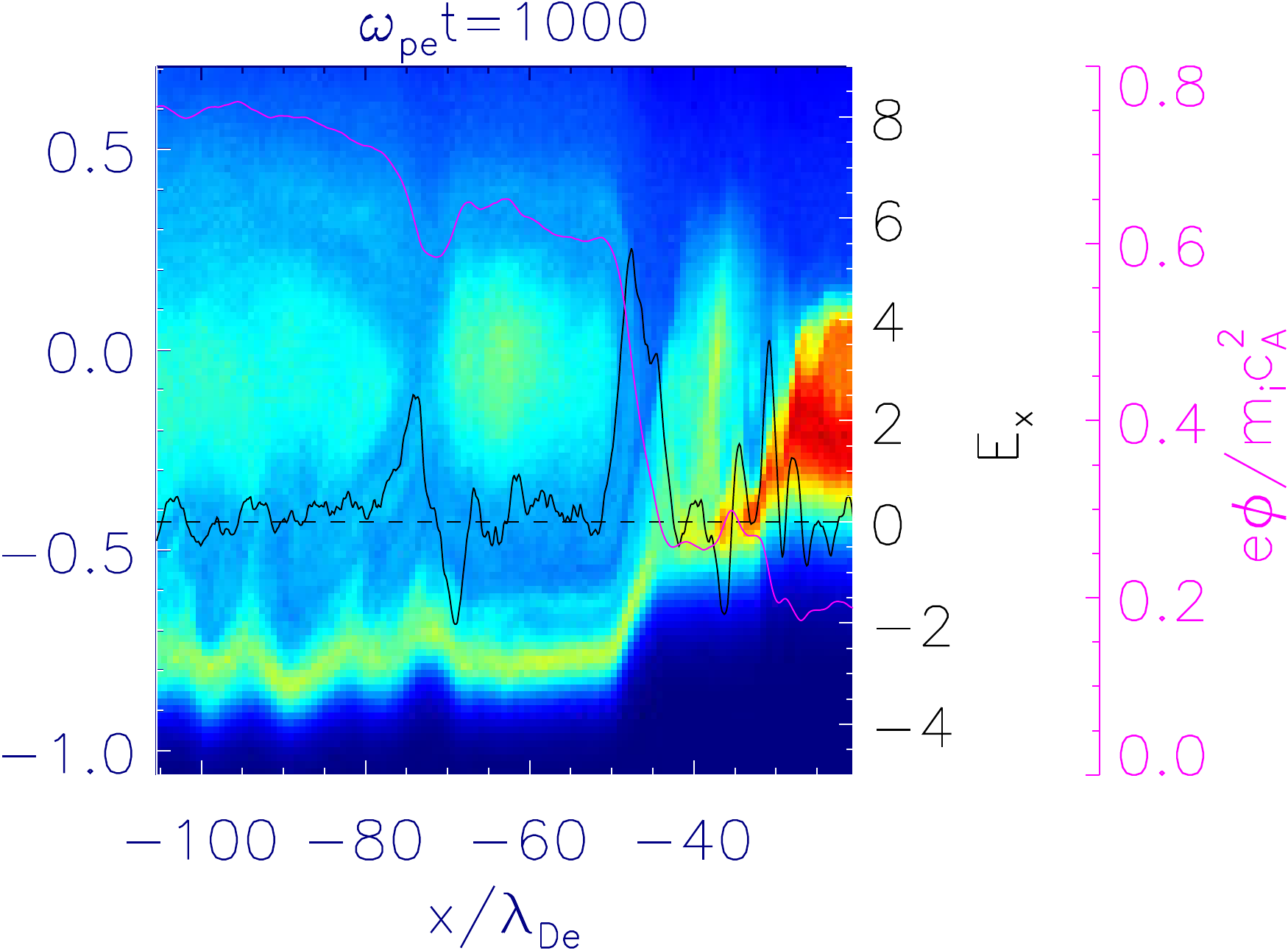}  
}
\vspace{0.2em}%
\\
\subfloat(c){
\includegraphics[scale=0.43]{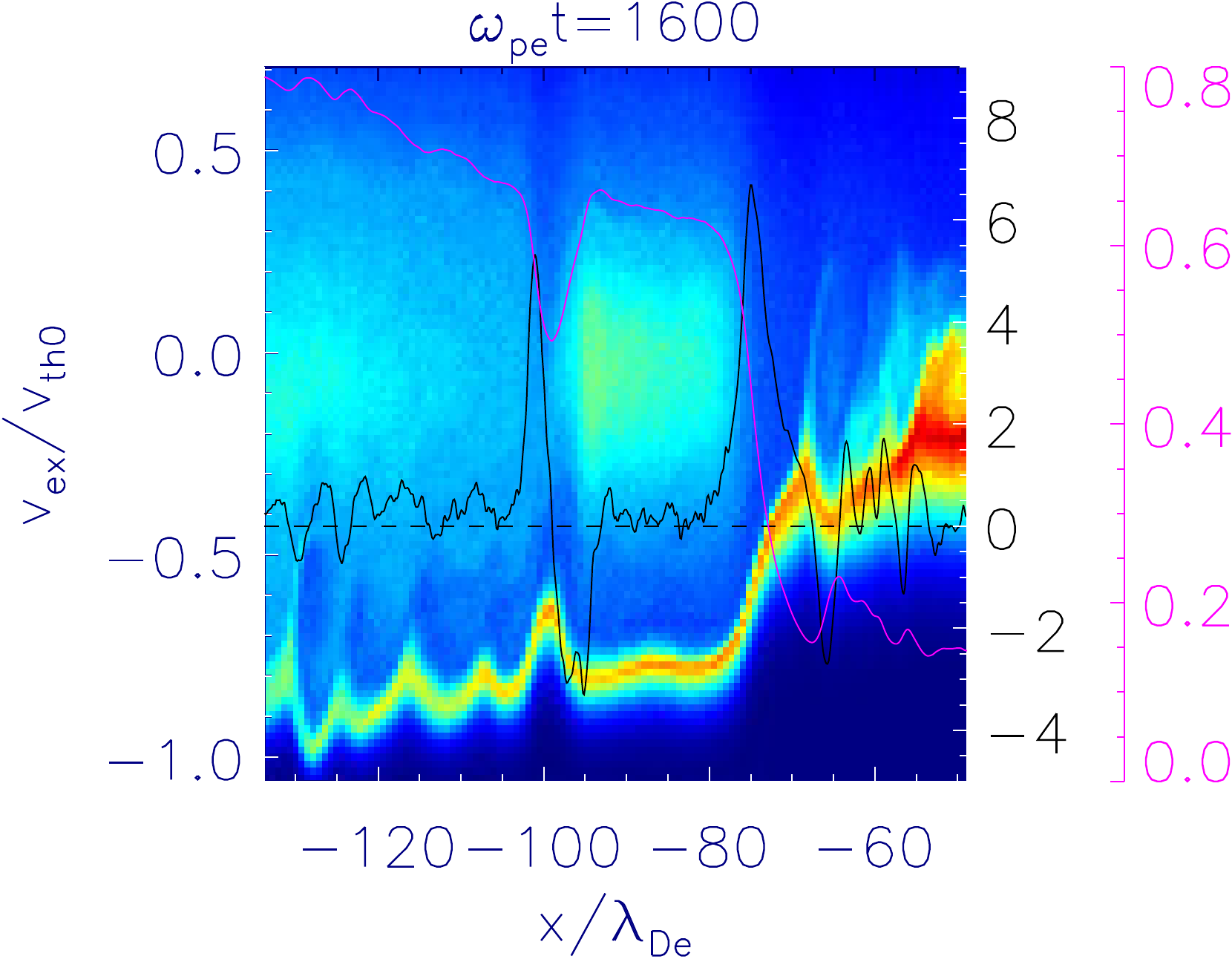}  
}
\vspace{-0.2em}%
\subfloat(d){
\includegraphics[scale=0.43]{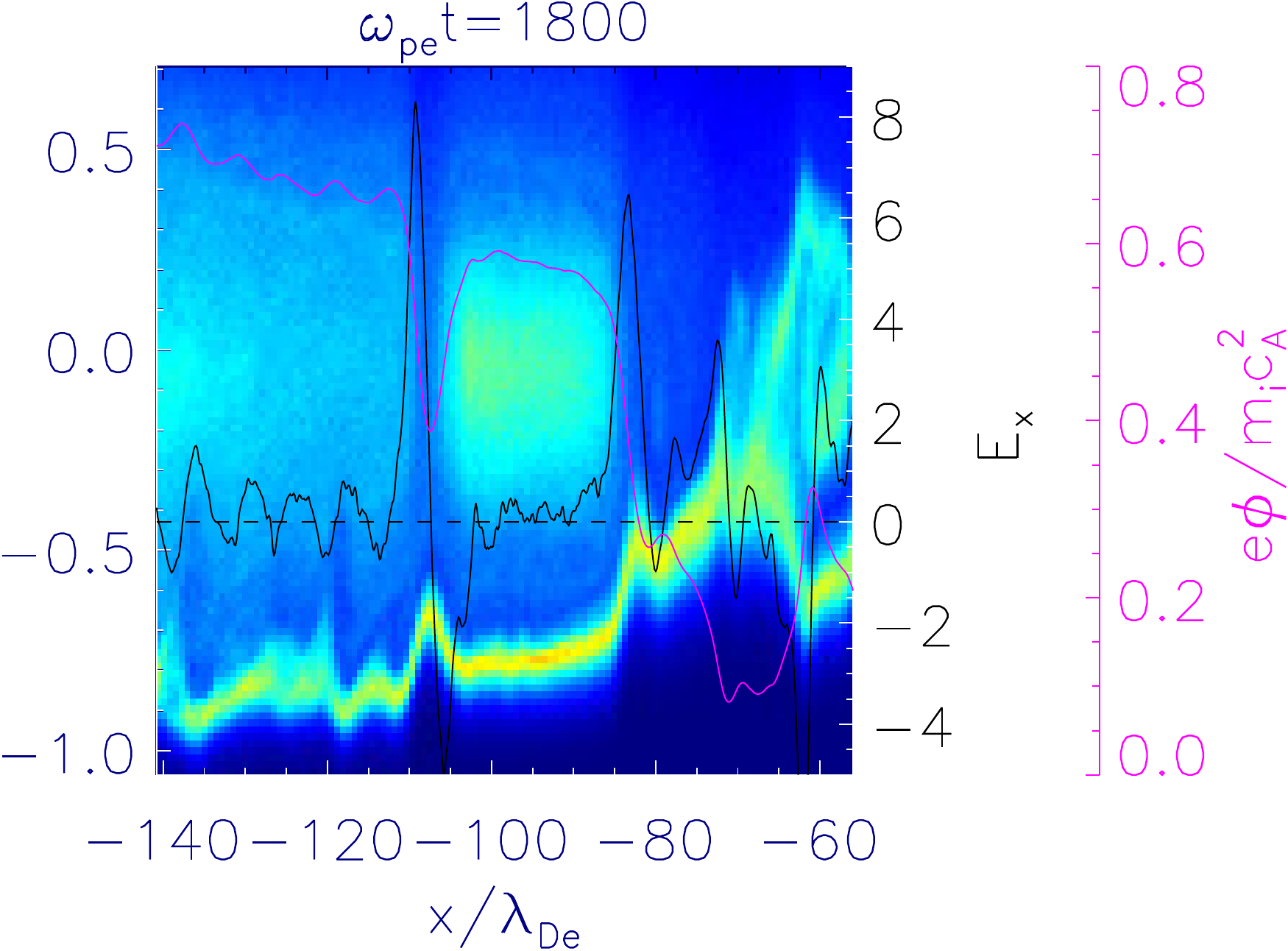}  
}
\vspace{-0.2em}%
\centering
\caption{ \label{etrap} Electron phase space at $\omega_{pe}t$=700 (a), 1000 (b), 1600 (c) and 1800 (d), showing electron trapping between two DLs. The same format is used as Figure \ref{ep_early}.     }
\end{figure*}

\subsection{\label{etrapping}Electron Trapping between DLs}

Another new dynamics observed in multi-DL systems is the trapping of electrons between DLs. We show in Figure \ref{etrap} the long time evolution of the electron phase space around two DLs. The electric field of both DLs grows over time. The second DL (on the left in (b) at $x\sim$ -75) is weaker than the first (on the right at $x\sim$ -45), i.e., it has a smaller drop in the potential than the first. Apparently, it is much more bipolar than an ideal monopolar DL as it develops a large negative $E_x$  (in (d)), which corresponds to a deep dip in the potential. This dip together with the potential drop at the first DL takes the form of a "cap" (an inverted potential well). Such a potential is capable of trapping electrons. The cap enlarges as the dip deepens from (a) to (d). With the following calculation, it can be demonstrated that the same (cyan) population of electrons stay within the cap during this entire period of $\sim$1000 $\omega_{pe}^{-1}$. These electrons are centered at approximately zero velocity with a velocity range of $\Delta v/v_{th0}\sim\pm$0.3 (in (d)). That means that their typical velocity is of order $\sim$0.3. Given the separation between the two DLs being $\Delta x\sim$ 20 $\lambda_{De}$, the transit time of the electrons across the separation is $\tau_{transit}$ = $\Delta x/\Delta v$ = 20/0.3 $\sim$ 70 $\omega_{pe}^{-1}$. This is much shorter than the 1000 $\omega_{pe}^{-1}$ time scale. Therefore, they are trapped between the two DLs. In order to trap these electrons, the potential cap needs to have a depth of $\Delta\phi/T_{h0}\sim\Delta v^2/v_{th0}^2\sim$ 0.3$^2\sim$ 0.1. In (d), taking the shallower left side, the cap has a depth of $e\phi_{cap}/T_{h0}\sim$ 0.2 (note $T_{h0}$=1), enough to trap these electrons. The trapped electrons are essentially a peak in phase space because of an enhancement in phase space density there. It has a spatial extent much wider than the typical DL width (of $\sim$ 10 $\lambda_{De}$) because the two DLs are widely separated. Given that electrons are trapped between DLs, there will not be any transport of energy across these regions. The more DLs develop, the more trapped regions may arise, and this provides inhibition of electron transport within the chain of DLs.

\section{\label{dis6}DISCUSSION AND SUMMARY}


 The impact of DLs on the ambient ions has important implications for understanding how hot electrons in the corona drive the solar wind. Hot electrons produced in the corona try to escape along the ambient magnetic field both toward the chromosphere and outward from the sun. In the process they drive vast numbers of DLs, which in turn accelerate the ambient ions, creating strong ion flows over large regions. Along open field lines, such ion flows move away from the the upper corona into the interplanetary space where they make up the solar wind. In the exospheric models of (fast) solar wind generation \citep{Lamy03}, an assumed interplanetary potential is required to accelerate ions to the solar wind speed. Such a potential is likely to be manifest as a chain of DLs. Indeed, based on DL measurements by the WIND spacecraft near 1 AU, it was estimated that a succession of DLs from the corona to the Earth could produce the total electric potential required by the exospheric models \citep{Lacombe02}.

We briefly compare some basic characteristics observed in the multi-DL systems in this work and those reported in previous 1D electrostatic PIC simulations driven by a strong applied potential \citep{Sato81}. In this earlier work it was found that the number of DLs formed increased with the system length. We also observe a similar tendency here. It was also shown that the total strength of the DLs became stronger with the system length as more DLs were formed, but was far weaker than the externally applied potential. In our case, the total DL strength $e\phi_{DL}$ increases moderately between the two smaller systems, indicated by the maxima of the black and orange curves in Figure \ref{4phi}. As the domain size further increases, $e\phi_{DL}$, however, appears to saturate (blue and red curves) and stays roughly at a constant value at late time. That value is comparable to the initial hot electron temperature. In Sato \& Okuda (1981), since $e\phi_{DL}$ was far less than the applied potential, which could be considered as the maximum available energy in the system, there was room for the DLs to grow stronger. In our case, however, $e\phi_{DL}$ already comes close to the initial temperature, which can similarly be understood as an upper limit on the potential. Further strengthening beyond this value is unlikely. This may be why more DLs in larger systems do not result in a stronger total potential jump here as they did in Sato \& Okuda (1981).


In conclusion, using large-scale PIC simulations, we reveal the self-consistent generation of multiple DLs during the transport of flare-heated electrons. Dynamics only observable in multi-DL systems are investigated for the first time. The primary DL enhances the return current, which is responsible for driving DLs, and results in subsequent breeding of secondary DLs. The chain of DLs occurs over an extended region. Electrons are trapped between DLs, suppressing the transport of electron heat in the trapped regions. The chain of DLs accelerates ambient ions, creating strong ion flows across the system. The simulations therefore reveal the mechanism by which hot electrons heated in the corona couple to ions to drive the strong outflows that make up the solar wind. This in turn separates adjacent DLs, expanding the DL breeding territory. These dynamics provide a more realistic picture of DL occurrence in the solar corona that is important for understanding energetic electron transport and in preparing for future space missions such as Solar Probe Plus.




\section*{Acknowledgments}
This work has been supported by NSF grants AGS-1202330 and ATN-0903964. The computations were performed at the National Energy Research Scientific Computing Center.




\begin{thebibliography}{39}
\expandafter\ifx\csname natexlab\endcsname\relax\def\natexlab#1{#1}\fi
\expandafter\ifx\csname href\endcsname\relax
  \def\href#1#2{}\fi
\expandafter\ifx\csname urllinklabel\endcsname\relax
  \def\urllinklabel{[LINK]}\fi
\expandafter\ifx\csname adsurllinklabel\endcsname\relax
  \def\adsurllinklabel{[ADS]}\fi

\bibitem[{{Angelopoulos}(2008)}]{Angelopoulos08}
{Angelopoulos}, V. 2008, \ssr, 141, 5


\bibitem[{Arber \& Melnikov(2009)}]{Arber09}
Arber, T.~D. \& Melnikov, V.~F. 2009, ApJ, 690, 238
 \href{http://stacks.iop.org/0004-637X/690/i=1/a=238}{\urllinklabel}

\bibitem[{{Aschwanden} {et~al.}(1996){Aschwanden}, {Kosugi}, {Hudson}, {Wills},
  \& {Schwartz}}]{Aschwanden96}
{Aschwanden}, M.~J., {Kosugi}, T., {Hudson}, H.~S., {Wills}, M.~J., \&
  {Schwartz}, R.~A. 1996, \apj, 470, 1198


\bibitem[{{Aschwanden} \& {Schwartz}(1995)}]{AschSchwartz95}
{Aschwanden}, M.~J. \& {Schwartz}, R.~A. 1995, \apj, 455, 699


\bibitem[{{Aschwanden} {et~al.}(1995){Aschwanden}, {Schwartz}, \&
  {Alt}}]{Asch95}
{Aschwanden}, M.~J., {Schwartz}, R.~A., \& {Alt}, D.~M. 1995, \apj, 447, 923


\bibitem[{{Battaglia} \& {Benz}(2006)}]{Battaglia06}
{Battaglia}, M. \& {Benz}, A.~O. 2006, A\&A, 456, 751


\bibitem[{Block(1978)}]{Block78}
Block, L.~P. 1978, Astrophys. Space Sci., 55, 59
 \href{http://dx.doi.org/10.1007/BF00642580}{\urllinklabel}

\bibitem[{{Ergun} {et~al.}(2009){Ergun}, {Andersson}, {Tao}, {Angelopoulos},
  {Bonnell}, {McFadden}, {Larson}, {Eriksson}, {Johansson}, {Cully}, {Newman},
  {Goldman}, {Roux}, {Lecontel}, {Glassmeier}, \& {Baumjohann}}]{Ergun09}
{Ergun}, R.~E., {Andersson}, L., {Tao}, J., {Angelopoulos}, V., {Bonnell}, J.,
  {McFadden}, J.~P., {Larson}, D.~E., {Eriksson}, S., {Johansson}, T., {Cully},
  C.~M., {Newman}, D.~N., {Goldman}, M.~V., {Roux}, A., {Lecontel}, O.,
  {Glassmeier}, K.-H., \& {Baumjohann}, W. 2009, Physical Review Letters, 102,
  155002


\bibitem[{{Ergun} {et~al.}(2001){Ergun}, {Su}, {Andersson}, {Carlson},
  {McFadden}, {Mozer}, {Newman}, {Goldman}, \& {Strangeway}}]{Ergun01}
{Ergun}, R.~E., {Su}, Y.-J., {Andersson}, L., {Carlson}, C.~W., {McFadden},
  J.~P., {Mozer}, F.~S., {Newman}, D.~L., {Goldman}, M.~V., \& {Strangeway},
  R.~J. 2001, Physical Review Letters, 87, 045003


\bibitem[{Guo(2010)}]{Guo10}
Guo, Y. 2010, Acta Astronautica, 67, 1063
 \href{http://www.sciencedirect.com/science/article/pii/S0094576510001980}{\urllinklabel}

\bibitem[{{Gurnett} \& {Pryor}(2012)}]{Gurnett12}
{Gurnett}, D.~A. \& {Pryor}, W.~R. 2012, Washington DC American Geophysical
  Union Geophysical Monograph Series, 197, 305


\bibitem[{{Hess} {et~al.}(2009){Hess}, {Mottez}, \& {Zarka}}]{Hess09}
{Hess}, S., {Mottez}, F., \& {Zarka}, P. 2009, \grl, 36, 14101


\bibitem[{{Krucker} {et~al.}(2010){Krucker}, {Hudson}, {Glesener}, {White},
  {Masuda}, {Wuelser}, \& {Lin}}]{Krucker10}
{Krucker}, S., {Hudson}, H.~S., {Glesener}, L., {White}, S.~M., {Masuda}, S.,
  {Wuelser}, J.-P., \& {Lin}, R.~P. 2010, ApJ, 714, 1108


\bibitem[{{Krucker} \& {Lin}(2008)}]{KruckerLin08}
{Krucker}, S. \& {Lin}, R.~P. 2008, ApJ, 673, 1181


\bibitem[{{Krucker} {et~al.}(2007){Krucker}, {White}, \& {Lin}}]{Krucker07}
{Krucker}, S., {White}, S.~M., \& {Lin}, R.~P. 2007, ApJ Lett., 669, L49


\bibitem[{{Lacombe} {et~al.}(2002){Lacombe}, {Salem}, {Mangeney}, {Hubert},
  {Perche}, {Bougeret}, {Kellogg}, \& {Bosqued}}]{Lacombe02}
{Lacombe}, C., {Salem}, C., {Mangeney}, A., {Hubert}, D., {Perche}, C.,
  {Bougeret}, J.-L., {Kellogg}, P.~J., \& {Bosqued}, J.-M. 2002, Annales
  Geophysicae, 20, 609


\bibitem[{Lamy {et~al.}(2003)Lamy, Pierrard, Maksimovic, \& Lemaire}]{Lamy03}
Lamy, H., Pierrard, V., Maksimovic, M., \& Lemaire, J.~F. 2003, Journal of
  Geophysical Research: Space Physics, 108, n/a
 \href{http://dx.doi.org/10.1029/2002JA009487}{\urllinklabel}

\bibitem[{{Levin} \& {Melnikov}(1993)}]{Levin93}
{Levin}, B.~N. \& {Melnikov}, V.~F. 1993, \solphys, 148, 325


\bibitem[{{Li} {et~al.}(2012){Li}, {Drake}, \& {Swisdak}}]{Li12}
{Li}, T.~C., {Drake}, J.~F., \& {Swisdak}, M. 2012, ApJ, 757, 20


\bibitem[{{Li} {et~al.}(2013){Li}, {Drake}, \& {Swisdak}}]{Li13}
---. 2013, ApJ, 778, 144
 \href{http://stacks.iop.org/0004-637X/778/i=2/a=144}{\urllinklabel}

\bibitem[{{Lin} {et~al.}(2002){Lin}, {Dennis}, {Hurford}, {Smith}, {Zehnder},
  {Harvey}, {Curtis}, {Pankow}, {Turin}, {Bester}, {Csillaghy}, {Lewis},
  {Madden}, {van Beek}, {Appleby}, {Raudorf}, {McTiernan}, {Ramaty}, {Schmahl},
  {Schwartz}, {Krucker}, {Abiad}, {Quinn}, {Berg}, {Hashii}, {Sterling},
  {Jackson}, {Pratt}, {Campbell}, {Malone}, {Landis}, {Barrington-Leigh},
  {Slassi-Sennou}, {Cork}, {Clark}, {Amato}, {Orwig}, {Boyle}, {Banks},
  {Shirey}, {Tolbert}, {Zarro}, {Snow}, {Thomsen}, {Henneck}, {McHedlishvili},
  {Ming}, {Fivian}, {Jordan}, {Wanner}, {Crubb}, {Preble}, {Matranga}, {Benz},
  {Hudson}, {Canfield}, {Holman}, {Crannell}, {Kosugi}, {Emslie}, {Vilmer},
  {Brown}, {Johns-Krull}, {Aschwanden}, {Metcalf}, \& {Conway}}]{Lin02}
{Lin}, R.~P., {Dennis}, B.~R., {Hurford}, G.~J., {Smith}, D.~M., {Zehnder}, A.,
  {Harvey}, P.~R., {Curtis}, D.~W., {Pankow}, D., {Turin}, P., {Bester}, M.,
  {Csillaghy}, A., {Lewis}, M., {Madden}, N., {van Beek}, H.~F., {Appleby}, M.,
  {Raudorf}, T., {McTiernan}, J., {Ramaty}, R., {Schmahl}, E., {Schwartz}, R.,
  {Krucker}, S., {Abiad}, R., {Quinn}, T., {Berg}, P., {Hashii}, M.,
  {Sterling}, R., {Jackson}, R., {Pratt}, R., {Campbell}, R.~D., {Malone}, D.,
  {Landis}, D., {Barrington-Leigh}, C.~P., {Slassi-Sennou}, S., {Cork}, C.,
  {Clark}, D., {Amato}, D., {Orwig}, L., {Boyle}, R., {Banks}, I.~S., {Shirey},
  K., {Tolbert}, A.~K., {Zarro}, D., {Snow}, F., {Thomsen}, K., {Henneck}, R.,
  {McHedlishvili}, A., {Ming}, P., {Fivian}, M., {Jordan}, J., {Wanner}, R.,
  {Crubb}, J., {Preble}, J., {Matranga}, M., {Benz}, A., {Hudson}, H.,
  {Canfield}, R.~C., {Holman}, G.~D., {Crannell}, C., {Kosugi}, T., {Emslie},
  A.~G., {Vilmer}, N., {Brown}, J.~C., {Johns-Krull}, C., {Aschwanden}, M.,
  {Metcalf}, T., \& {Conway}, A. 2002, \solphys, 210, 3


\bibitem[{{Lin} {et~al.}(2003){Lin}, {Krucker}, {Hurford}, {Smith}, {Hudson},
  {Holman}, {Schwartz}, {Dennis}, {Share}, {Murphy}, {Emslie}, {Johns-Krull},
  \& {Vilmer}}]{Lin03}
{Lin}, R.~P., {Krucker}, S., {Hurford}, G.~J., {Smith}, D.~M., {Hudson}, H.~S.,
  {Holman}, G.~D., {Schwartz}, R.~A., {Dennis}, B.~R., {Share}, G.~H.,
  {Murphy}, R.~J., {Emslie}, A.~G., {Johns-Krull}, C., \& {Vilmer}, N. 2003,
  ApJ Lett., 595, L69


\bibitem[{{Mangeney} {et~al.}(1999){Mangeney}, {Salem}, {Lacombe}, {Bougeret},
  {Perche}, {Manning}, {Kellogg}, {Goetz}, {Monson}, \& {Bosqued}}]{Mangeney99}
{Mangeney}, A., {Salem}, C., {Lacombe}, C., {Bougeret}, J.-L., {Perche}, C.,
  {Manning}, R., {Kellogg}, P.~J., {Goetz}, K., {Monson}, S.~J., \& {Bosqued},
  J.-M. 1999, Annales Geophysicae, 17, 307


\bibitem[{{Manheimer}(1977)}]{Manheimer77}
{Manheimer}, W.~M. 1977, Physics of Fluids, 20, 265


\bibitem[{Masuda {et~al.}(1994)Masuda, Kosugi, Hara, Tsuneta, \&
  Ogawara}]{Masuda94}
Masuda, S., Kosugi, T., Hara, H., Tsuneta, S., \& Ogawara, Y. 1994, Nature,
  371, 495


\bibitem[{{Masuda} {et~al.}(2000){Masuda}, {Sato}, {Kosugi}, \&
  {Sakao}}]{Masuda00}
{Masuda}, S., {Sato}, J., {Kosugi}, T., \& {Sakao}, T. 2000, Advances in Space
  Research, 26, 493


\bibitem[{{McKean} {et~al.}(1990){McKean}, {Winglee}, \& {Dulk}}]{McKean90}
{McKean}, M.~E., {Winglee}, R.~M., \& {Dulk}, G.~A. 1990, ApJ, 364, 295


\bibitem[{Mozer {et~al.}(2013)Mozer, Bale, Bonnell, Chaston, Roth, \&
  Wygant}]{Mozer13}
Mozer, F., Bale, S., Bonnell, J., Chaston, C., Roth, I., \& Wygant, J. 2013,
  Physical Review Letters, 111, 235002


\bibitem[{{Mozer} {et~al.}(1985){Mozer}, {Boehm}, {Cattell}, {Temerin}, \&
  {Wygant}}]{Mozer85}
{Mozer}, F.~S., {Boehm}, M.~H., {Cattell}, C.~A., {Temerin}, M., \& {Wygant},
  J.~R. 1985, Space Science Reviews, 42, 313


\bibitem[{Oreshina \& Somov(2011)}]{Oreshina11}
Oreshina, A. \& Somov, B. 2011, Moscow University Physics Bulletin, 66, 286
 \href{http://dx.doi.org/10.3103/S0027134911030167}{\urllinklabel}

\bibitem[{Raadu \& Rasmussen(1988)}]{Raadu88}
Raadu, M.~A. \& Rasmussen, J.~J. 1988, Astrophys. Space Sci., 144, 43
 \href{http://dx.doi.org/10.1007/BF00793172}{\urllinklabel}

\bibitem[{{Sato} \& {Okuda}(1981)}]{Sato81}
{Sato}, T. \& {Okuda}, H. 1981, J. Geophys. Res., 86, 3357


\bibitem[{{Sim{\~o}es} \& {Kontar}(2013)}]{Simoes13}
{Sim{\~o}es}, P.~J.~A. \& {Kontar}, E.~P. 2013, A\&A, 551, A135


\bibitem[{{Smith} \& {Lilliequist}(1979)}]{Smith79}
{Smith}, D.~F. \& {Lilliequist}, C.~G. 1979, ApJ, 232, 582


\bibitem[{{Spitzer}(1962)}]{Spitzer62}
{Spitzer}, L. 1962, {Physics of Fully Ionized Gases}


\bibitem[{{Tomczak}(2009)}]{Tomczak09}
{Tomczak}, M. 2009, \aap, 502, 665


\bibitem[{{Tsuneta} {et~al.}(1997){Tsuneta}, {Masuda}, {Kosugi}, \&
  {Sato}}]{Tsuneta97}
{Tsuneta}, S., {Masuda}, S., {Kosugi}, T., \& {Sato}, J. 1997, \apj, 478, 787


\bibitem[{{Tsytovich}(1971)}]{Tsytovich71}
{Tsytovich}, V.~N. 1971, Plasma Physics, 13, 741


\bibitem[{{Zeiler} {et~al.}(2002){Zeiler}, {Biskamp}, {Drake}, {Rogers},
  {Shay}, \& {Scholer}}]{Zeiler02}
{Zeiler}, A., {Biskamp}, D., {Drake}, J.~F., {Rogers}, B.~N., {Shay}, M.~A., \&
  {Scholer}, M. 2002, J. Geophys. Res., 107, 1230


\end{thebibliography}



\end{document}